\documentclass[twocolumn]{revtex4}
\usepackage{graphicx}

\begin{document}

\title{Resonant levels, vacancies and doping in Bi$_2$Te$_3$, Bi$_2$Te$_2$Se and Bi$_2$Se$_3$ tetradymites}

\author{Bartlomiej Wiendlocha}
\email[email: ]{wiendlocha@fis.agh.edu.pl}
\affiliation{AGH University of Science and Technology, Faculty of Physics and Applied Computer Science, Al. Mickiewicza 30, 30-059 Krakow, Poland}

\date{\today}

\begin{abstract}
Electronic structures of the tetradymites, Bi$_2$Te$_3$, Bi$_2$Te$_2$Se and Bi$_2$Se$_3$, containing various dopants and vacancies, are studied using the first principles calculations methods. We focus on the possibility of formation of the resonant levels (RL), confirming the formation of the RL by Sn in Bi$_2$Te$_3$, and predicting similar behavior of Sn in Bi$_2$Te$_2$Se and Bi$_2$Se$_3$. Vacancies, which are likely present on the chalcogen atoms in the real samples of Bi$_2$Te$_2$Se and Bi$_2$Se$_3$, are also studied and their charged donor and resonant behavior is discussed. 
Doping of the vacancy-containing materials with regular acceptors, like Ca or Mg, is shown to compensate the donor effect of vacancies, and $n-p$ crossover, while increasing the dopant concentration, is observed. 
We verify, that RL on Sn is not disturbed by the chalcogen vacancies in Bi$_2$Te$_2$Se and Bi$_2$Se$_3$, and 
for the Sn-doped materials with Se or Te vacancies, double-doping, instead of heavy doping with Sn, is suggested as an effective way of reaching the resonant level. This should help to avoid the smearing of the RL, which was a possible reason for an earlier unsuccessful experimental observation of the influence of the RL on thermoelectric properties of Sn doped Bi$_2$Te$_2$Se. 
Finally we show, that Al and Ga are possible new resonant impurities in the tetradymites, hoping that it will stimulate further experimental studies.
\end{abstract}

\maketitle

\section{Introduction}
\label{intro}

The tetradymites are a group of minerals of the general formula (Bi,Sb)$_2$(Te,Se)$_3$, very intensively studied as the thermoelectric materials operating near room temperature, for Peltier cooling { or power generation} applications,~\cite{snyder-nmat} and, more recently, as the topological insulators (TI)~\cite{cava-tetra_cryst,topol-rmp,cava-bi2se3-prb09}.
The best known material from this family is Bi$_2$Te$_3$, which alloys have a large thermoelectric figure of merit, ZT, around 1.0 near room temperatures,~\cite{snyder-nmat} and the best tetradymites alloys and nanocomposites reach even ZT = 1.5 around $T = 400$~K~\cite{goldsmid-rev2014}.
In this work we focus on the electronic structure of the bismuth-based tetradymites, Bi$_2$Te$_3$, Bi$_2$Te$_2$Se, and Bi$_2$Se$_3$, containing impurity atoms and vacancies. 
Band structure of the pure compounds is quite well known, both theoretically and experimentally (see, Refs. ~\cite{Caywood1970,kohler1974,Pecheur1989,jepsen1997,mahanti2002,Greanya2002,Kioupakis2010,topol-first_princ,bi2se3-direct} and references therein). However, much less theoretical work was devoted to the electronic structure of materials with defects or impurities, especially if compared e.g. to the lead telluride family, which is likely related to the more complicated crystal structure of the tetradymites. This was one of the motivations for this work, and our aim was to systematically study the main features of the electronic structure of tetradymites containing various dopant atoms and vacancies. 
We focus on the so-called resonant impurities, starting with already experimentally investigated Sn in Bi$_2$Te$_3$, and extending those studies on remaining two materials. 
The second aspect of present work is to shed some more light on the vacancy behavior in Bi$_2$Se$_3$ and Bi$_2$Te$_2$Se, and its effect on the possibility of the RL formation. The chemistry of defects in Bi$_2$Te$_3$ is much better known and controlled~\cite{scherrer}, thus we discuss the two other materials, which are naturally $n$-type, and vacancies~\cite{cava-tetra_cryst} (either on Se or Te) are often considered as a reason for this. 
We explore, whether the presence of vacancies may alter the formation of the resonant state on Sn, and verify whether double doping with the already known effective $p$-type impurities (Ca and Mg) may be an efficient way of tuning the Fermi level position in Sn doped material, alternative to heavy doping with Sn. 
Finally, we propose two new, potentially interesting impurities, which may form resonant levels in tetradymites: Al and Ga, to stimulate further experimental works in this family of materials.

\section{Computational details}

All the three studied compounds, Bi$_2$Te$_3$, Bi$_2$Te$_2$Se, and Bi$_2$Se$_3$, crystallize in the rhombohedral structure,~\cite{cava-tetra_cryst,bi2te3-cryst} space group no. 166, $R$-$3m$, consisting of a sequence of atomic layers, Te(1) - Bi - Te(2) - Bi - Te(1) (Bi$_2$Te$_3$), Te(1) - Bi - Se(2) - Bi - Te(1) (Bi$_2$Te$_2$Se) and finally,  Se(1) - Bi - Se(2) - Bi - Se(1) (Bi$_2$Se$_3$). This sequence of five layers is often called a quintuple layer, atoms within the layer are covalently bounded, whereas between the quintuple layers there are weaker van der Waals bonds~\cite{cava-tetra_cryst,topol-first_princ}.
Experimental unit cell parameters and atomic positions, for each of the structure, are gathered in Table~\ref{tab:cryst}, and those values were used in calculations performed in this work. In the unit cell of tetradymites, atoms occupy three nonequivalent crystal sites. Bi and Te(1)/Se(1) are located at (2c) sites, with coordinates $(x,x,x)$, whereas Te(2)/Se(2) atoms are placed at (1a) i.e. (0,0,0) position.

Electronic structure calculations were done within the density functional theory (DFT) formalism. The Korringa-Kohn-Rostoker (KKR) method was used~\cite{kaprzyk90,kkr99,stopa2004} in the semirelativistic and spherical potential approximations, which are sufficient to describe the densities of states or the formation of the resonant levels.
The chemical disorder, caused by the presence of the impurity atoms or vacancies, was simulated using the coherent potential approximation (CPA)~\cite{kaprzyk90,kkr99,ebert-kkr2011}. In this approach, using the Green function techniques, the real disordered system (e.g. binary alloy A$_x$B$_{1-x}$) is replaced by the ordered system of effective ''CPA atoms'', described by the effective Green function. This effective Green function is calculated self-consistenlty, employing the CPA condition, in which replacing one of the ''CPA atoms'' by component A or B atom does not lead to an additional scattering and do not change the averaged properties of the medium. CPA allows to consider very small impurity concentrations, as small as $x = 0.1\%$, and all calculations for $0 \leq x \leq 1$ use the same primitive cell of the host material. Computations in a single unit cell does not 
allow, however, to study the crystal structure relaxation process around impurities or vacancies, thus any relaxation effects are neglected in our computations.
In order to obtain the high accuracy of the Fermi level position in the doped materials, the Lloyd formula was used~\cite{kaprzyk90}.
Local density approximation (LDA) was used, with the parametrization of von Barth and Hedin~\cite{lda}.
Very dense ${\bf k}$-point meshes for the self consistent cycle and density of states calculations were used, up to $\sim 4600$ points in the irreducible part of the Brillouin zone. 
To increase the unit cell filling, empty spheres with the atomic number $Z = 0$ were added between each of the atomic layers.

{
For the selected case of Sn doped Bi$_2$Te$_3$, additional calculations were performed, using the supercell technique and the full potential linearized augmented plane wave method (FP-LAPW, the WIEN2k code\cite{wien2k}). This was done to verify whether effects, neglected within the KKR-CPA approach (non-sperical potential, crystal lattice relaxation after Sn/Bi atomic substitution, or the spin-orbit coupling), may wash out the resonant state. As the answer was negative and RL }{ formation was confirmed in relaxed FP-LAPW + spin-orbit case (see, below), we may expect
that the main conclusions of this work, related to the presence of RL, do not depend on the crystal structure relaxation or spin-orbit interaction.}

\begin{table}[t]
\caption{\ Experimental~\cite{cava-tetra_cryst,topol-first_princ,bi2te3-cryst} unit cell parameters of Bi$_2$Te$_3$, Bi$_2$Te$_2$Se and Bi$_2$Se$_3$ in the rhombohedral convention of the unit cell: lattice constant, $a$ (\AA), and rhombohedral angle, $\alpha_{\rm Rh}$ ($^{\circ}$)). Space group no. 166, $R$-3$m$, $x$ is the parameter of the (2c) $(x,x,x)$ site, occupied by Bi and Te(1)/Se(1), Te(2)/Se(2) atoms occupy (1a) site, with coordinates (0,0,0).}
\label{tab:cryst}
\begin{tabular}{llll}
\hline\noalign{\smallskip}
    \hline
    Parameter & Bi$_2$Te$_3$ & Bi$_2$Te$_2$Se & Bi$_2$Se$_3$ \\
    \noalign{\smallskip}\hline\noalign{\smallskip}
    $a$ (\AA) & 10.473 & 10.230 & 9.841\\
    $\alpha_{\rm Rh}$ ($^{\circ}$) & 24.16 & 24.25 & 24.27\\
    x Bi  & 0.400 & 0.3958 & 0.399\\
    x Te(1)/Se(1)  & 0.2095 & 0.2118 & 0.206\\
    x Te(2)/Se(2)  & 0 & 0 & 0\\
    \noalign{\smallskip}\hline
  \end{tabular}
\end{table}

\section{Results and discussion}\label{sec:results}

\subsection{Sn resonant level\label{sec:sn_te}}

\begin{figure}[t]
\includegraphics[width=0.50\textwidth]{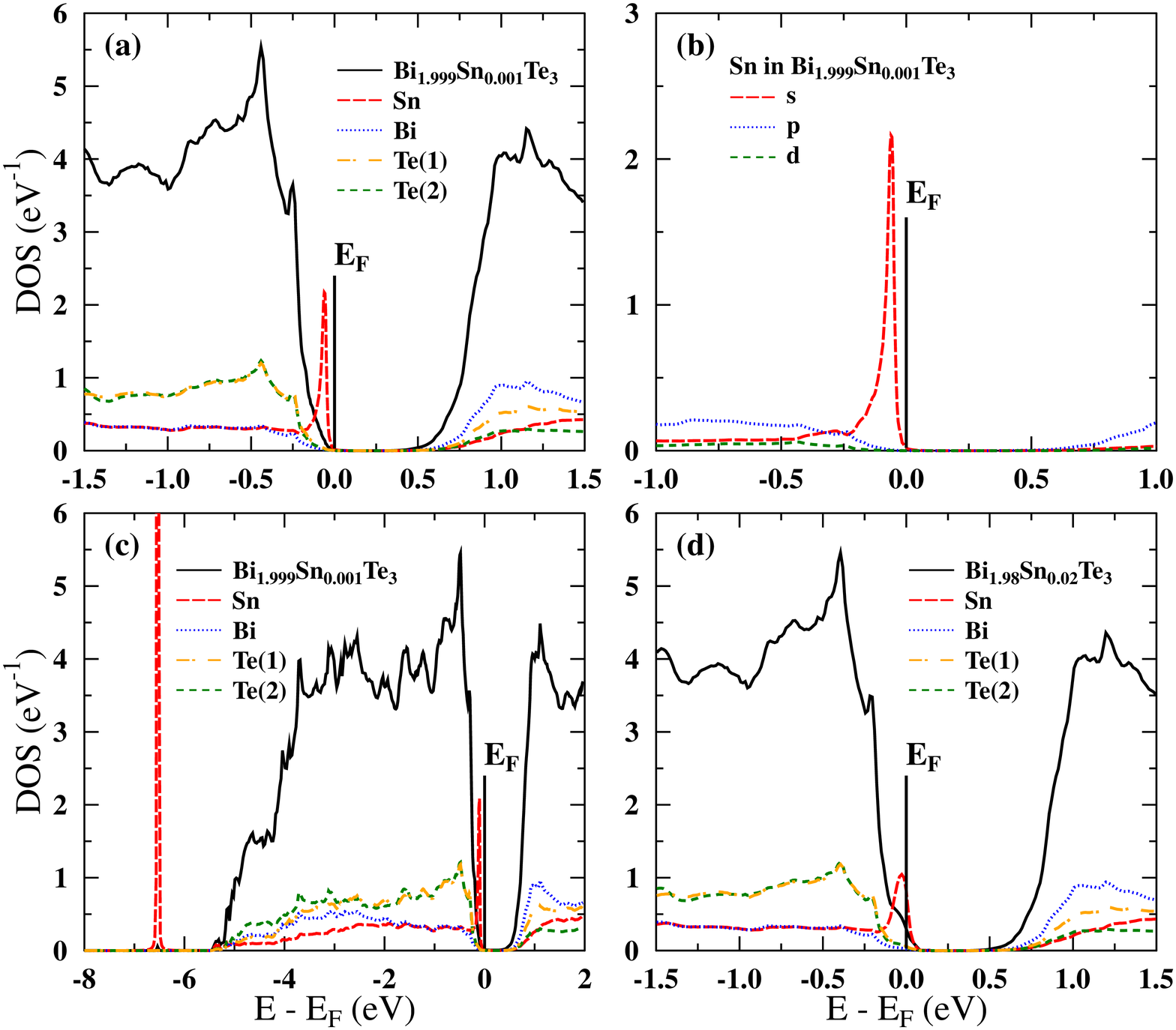}
  \caption{Calculated DOS of Bi$_{2-x}$Sn$_x$Te$_3$, (a)-(c): $x = 0.001$, (d): $x = 0.02$. Panels (a) and (d) show the total DOS for $E = E_F \pm 1.5$ eV; panel (b) shows partial Sn DOS with angular momentum decomposition; panel (c) shows DOS in a broader energy range to show the hyper-deep DOS peak of Sn, around -6.5 eV below $E_F$. For all the panels and all other figures in this work, total DOS (solid black line) is given per formula unit, whereas partial atomic densities of states, plotted with colors, are given per single atom, not multiplied by its concentration.}
  \label{fig:te}
\end{figure}

\begin{figure}[t]
\includegraphics[width=0.50\textwidth]{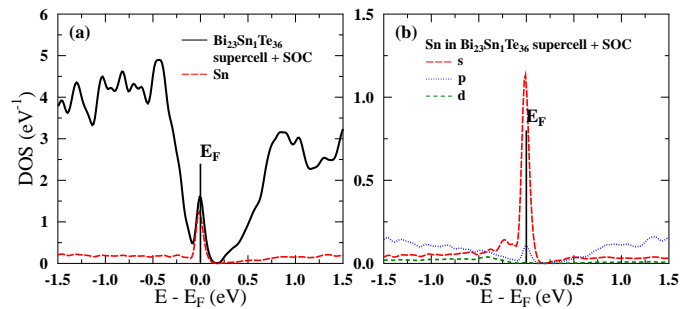}
  \caption{DOS of the relaxed Bi$_{23}$Sn$_1$Te$_{36}$ supercell, computed with spin-orbit coupling taken into account. (a) black solid line: total DOS, divided by 12 (the number of Bi$_2$Te$_3$ formula units in the supercell), red dashed line: Sn atomic DOS; (b) angular momentum decomposition of the Sn atomic DOS. RL is not removed by either the crystal structure relaxation, or the spin-orbit interaction.}
  \label{fig:supercell}
\end{figure}

We start our discussion with the very intriguing acceptor in Bi$_2$Te$_3$, which is tin. 
Sn was found to be the so-called resonant impurity in Bi$_2$Te$_3$~\cite{bi2te3-sn-1988,bi2te3-sn-1998,bi2te3-sn-heremans} which is a special case of an impurity behavior in semiconductors. From the band structure calculations point of view, fingerprint of such a resonant level, called also a virtual bound state~\cite{friedel,ees-review,cpa-gyorffy1}, is a narrow peak of the density of electronic states (DOS) of the impurity atom at the resonance energy, when the concentration of the impurity atoms is very small (here we use 0.1\% to indentificate the RL formation). Due to the presence of such a peak, the doped system, containing resonant impurities, does not follow the rigid band model~\cite{bw2013,bw2014-apl,bw2014-jap,BW2015},  where the doping would lead to the rigid shift of the Fermi level only.
RL can have various effects on the electronic and transport properties of the material, depending on its details, like the position versus the band edges, the orbital momentum quantum number or degree of hybridization with the host crystal states. To list several examples, RL may be a charge trap, when it is in the gap of the semiconductor (e.g. Indium in PbTe at room temperature~\cite{ees-review}); a strong resonant scattering center, as in the case of transition metal - noble metals alloys~\cite{cpa-gyorffy1};
may form the defect states which pin the Fermi level and lead to the partial charge localization (like 3d states of Ti in the conduction band of PbTe~\cite{bw2014-apl,konig-pbte_ti}). All these possibilities can be very interesting subjects for various physical and chemical studies.
From the point of view of the thermoelectric materials based on semiconductors, the most interesting is the case where the RL is close to the band edge (and thus near the Fermi level) and (partly) hybridize with the host electronic states, creating a redistribution of the electronic states around a valence (or conduction) band, which may lead to the enhancement of the thermopower ({\it S}) and thermoelectric power factor (PF).
Such an example is PbTe doped with Tl~\cite{heremans-science}, where the presence of RL leads to an increase in the thermopower (from $50-55$ $\mu$V/K to $120-140$~$\mu$V/K) around the carrier concentrations of $p \simeq 5-10\times 10^{19}$ cm$^{-3}$~\cite{heremans-science,ees-jaworski}. This increase in the thermopower may be understood as the effect of the distortion and increase in the density of electronic states near $E_F$, created by the resonant peak of DOS~\cite{ees-review,heremans-science}, or, as was recently shown~\cite{bw2013}, compared to the effect of the increase in the band degeneracy, since the presence of the RL leads to the increase of the number of electronic states available around the valence band, without forming the isolated impurity band~\cite{bw2013}.

\begin{figure}[t]
\includegraphics[width=0.50\textwidth]{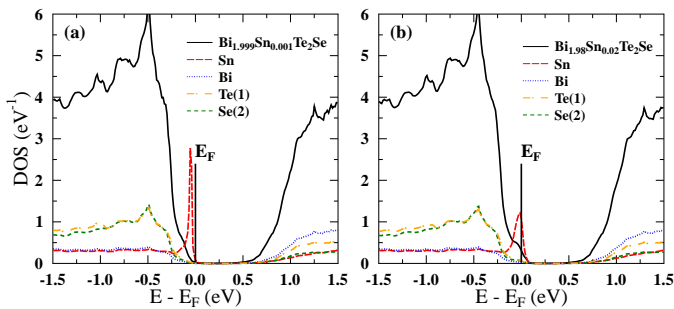}
  \caption{DOS of Bi$_{2-x}$Sn$_x$Te$_2$Se, (a): $x = 0.001$, (b): $x = 0.02$. The RL DOS peak at Sn atom is clearly visible.}
  \label{fig:tese}
\end{figure}

\begin{figure}[b]
  \includegraphics[width=0.50\textwidth]{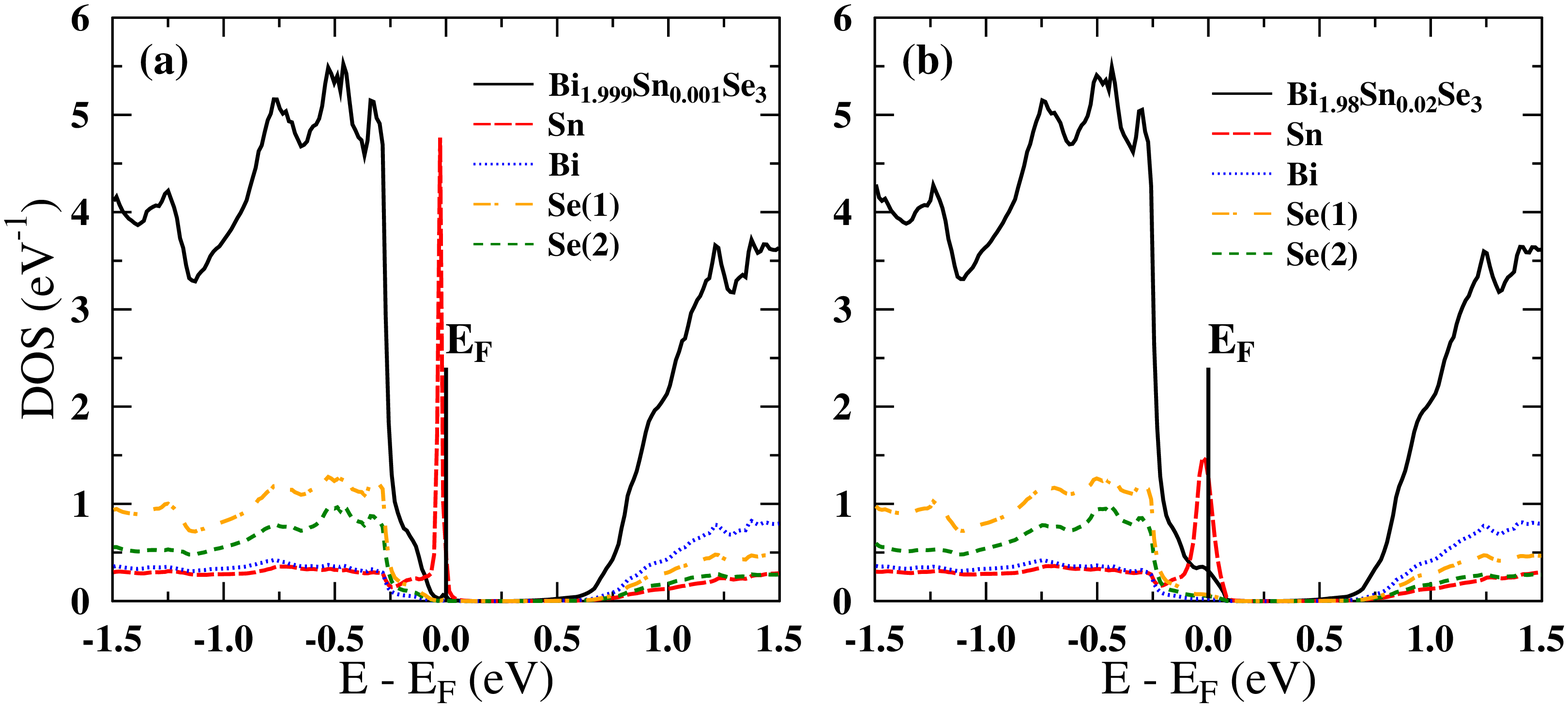}
  \caption{DOS of Bi$_{2-x}$Sn$_x$Se$_3$, (a): $x = 0.001$, (b): $x = 0.02$. The RL DOS peak at Sn atom is clearly visible. }
  \label{fig:se}
\end{figure}

Other examples of successful realization of the idea of improving the thermoelectric efficiency using the RL are PbTe:Tl alloyed with Si~\cite{pbte-tl-si-nano}, Se and S~\cite{ees-jaworski}, SnTe:In~\cite{snte_in-ren}, or Bi$_2$Te$_3$:Sn~\cite{bi2te3-sn-heremans}, on which we are focusing now.

In $p$-type Bi$_{2-x}$Sn$_x$Te$_3$, for $x \simeq 1\% - 2\%$ ($p \simeq 3 - 6\times 10^{19}$ cm$^{-3}$) at room temperature, thermopower was found to be about two times larger ($210 - 220$~$\mu$V/K) than achieved at the same carrier concentrations but using different acceptors ($100 - 140$~$\mu$V/K for Pb- or Ge-doped samples). This increase, basing also on earlier studies (see, Refs. ~\cite{bi2te3-sn-1988,bi2te3-sn-1998,bi2te3-sn-heremans} and references therein)  was explained as due to the formation of the resonant level by Sn atoms. 
For larger Sn concentrations, $x = 5\%$, the efficiency of RL was diminished and $S$ was brought back close to the Pisarenko relation ($S$ versus carrier concentration) for a ''regular'' Bi$_2$Te$_3$. Such a lost in RL efficiency for larger Sn concentrations may be related to two factors. First is the possible departure of $E_F$ from the optimal position that maximized the thermopower: as Sn dopes the system $p$-type, it also controls the position of $E_F$. Second is smearing of resonant states, as the impurity concentration increases, the level of hybridization with the host band structure becomes larger and RL broadens, gradually loosing the ''resonant'' character, being integrated into the host electronic structure. This also reduces the height of the DOS peak, connected to the RL, as will be presented below.

Our KKR-CPA electronic structure calculations confirm the formation of RL by Sn in Bi$_2$Te$_3$, as presented in 
Fig.~\ref{fig:te}. Figures~\ref{fig:te}(a)-(c) show the DOS for $x = 0.001$ doping case. We clearly see, that Sn forms a sharp, resonant DOS peak near the edge of the valence band (and $E_F$). In Fig.~\ref{fig:te}(b), the partial Sn DOS is plotted, with the angular momentum decomposition. The orbital momentum character of this DOS peak is mainly s-like, thus the RL is formed by the 5s orbitals of tin. 
Sn has a double role in Bi$_2$Te$_3$, as it also dopes the system $p$-type: when the concentration of Sn atoms increases, $E_F$ is moving deeper into the valence band, Fig.~\ref{fig:te}(d). At the same time, the resonant peak broadens and hybridizes with the valence band DOS. 
Figure~\ref{fig:te}(c) shows the DOS for the larger energy range, and we see that Sn in Bi$_2$Te$_3$ creates two resonant levels, one at the valence band edge, and the second one, so-called hyper-deep defect state (HDS)~\cite{Hjalmarson1980} at -6.5 eV below $E_F$ (in this terminology the RL near the valence band edge is called deep defect state, DDS). This state is also s-like and these two resonant states usually appear together, as the analog of bonding and anti-bonding pair of orbitals. Similar situation was found for the group III impurities (Tl, In, Ga) in PbTe~\cite{mahanti-prl06,mahanti-prb08}.
It is worth mentioning here, that recently, the formation of sole HDS (without DDS near $E_F$) was found to be a reason for the acceptor behavior of In, Ga and Sn in elemental bismuth, in spite of isovalent character of In and Ga with Bi, and this was proposed to be a novel doping mechanism in solids~\cite{Jin2015}.

As was mentioned before, to verify whether effects, neglected within the KKR-CPA approach (non-spherical potential, crystal lattice relaxation after Sn/Bi atomic substitution, or the spin-orbit coupling), may wash out the resonant state, complementary FP-LAPW computations, using the WIEN2k code~\cite{wien2k}, were done. The rhombohedral primitive cell was 
transformed into the hexagonal equivalent, containing 3 formula units (15 atoms), and the hexagonal unit cell was multiplied 2x2x1 times, forming the 60-atom hexagonal supercell, containing 12 Bi$_2$Te$_3$ formula units. Single Bi atom was then replaced by Sn, and resulting supercell had the $P$3$m$1 space group (no. 156).
The Bi$_{23}$Sn$_1$Te$_{36}$  supercell was next optimized, to investigate the lattice relaxation around Sn impurity, in the semirelativistic approach, and using the Perdew-Wang LDA~\cite{pw91}. We have found only a small negative chemical pressure effect, which led to the decrease of the nearest- and next-nearest-neighbors distance around Sn (-1.6\% and -0.5\%, respectively). This is similar to what was observed in Sn doped Bi~\cite{Jin2015}. For the relaxed supercell, the computations including spin-orbit coupling (SOC) were done, on a 10x10x3 k-point mesh. The resulting densities of states are presented in Fig.~\ref{fig:supercell}, and we see, that the formation of RL is generally not affected, either by the relaxation process, or by the spin-orbit coupling. This is partially due to the s-like nature of the resonance, confirmed in Fig.~\ref{fig:supercell}(b). 
Although one has to remember that the details of the band structure of the tetradymites do depend on the spin-orbit coupling (especially the topological states), the larger-energy-scale effects, including the formation of s-like RL, depend on SOC much weaker. The exact position of RL versus the band edge may be altered by SOC, but the qualitative conclusion on the existence of RL occurred to be independent of SOC. Similar conclusion was earlier established for the Tl doped PbTe~\cite{bw2013,mahanti-prb08}, and in the remaining part of this work we rely on the semirelativistic KKR-CPA results.

Having confirmed the formation of the RL in Bi$_2$Te$_3$ by Sn we may now verify the resonant versus classic (rigid-band like) behavior of Sn and other impurities in the tetradymite series of compounds. 
In Fig.~\ref{fig:tese} and Fig.~\ref{fig:se} we show DOS of Sn doped Bi$_2$Te$_2$Se and Bi$_2$Se$_3$, respectively. Similarly to the Bi$_2$Te$_3$ case, Sn forms the resonant peak, which becomes more narrow and larger when Se substitutes Te. As in the previous case, peak is $s$-like and accompanied by a hyperdeep state, below the valence band (not shown here). Fig.~\ref{fig:tese}(b) and \ref{fig:se}(b), where $x = 0.02$ show, that also here Sn behaves as an acceptor, effectively doping the system $p$-type. Thus, according to our DFT-LDA calculations, Sn should be the resonant impurity also in Bi$_2$Te$_2$Se and Bi$_2$Se$_3$.

The literature concerning experimental studies on Bi$_2$Se$_3$:Sn is very limited, to the best author's knowledge there is only one report, which classifies Sn as an acceptor~\cite{bi2se3-sn-old}. Some more experimental data can be found on Bi$_2$Te$_2$Se:Sn, where Sn was also found to be an acceptor~\cite{bi2te2se-sn-prb,bi2te2se-sn-fuccillo,bi2te2se-sn-cava1},  
but no significant enhancement in the thermopower due to Sn doping was found~\cite{bi2te2se-sn-fuccillo}.
However, the situation in these two sister compounds is more complicated, comparing to Bi$_2$Te$_3$ case, due to the inseparable presence of large amount of defects, which make both compounds naturally $n$-type~\cite{cava-tetra_cryst}.

Thermoelectric properties of a series of Sn doped Bi$_{2}$Te$_2$Se samples were invesrigated in Ref.~\cite{bi2te2se-sn-fuccillo}, and it was suggested that the presence of 0.5\% of Se vacancies in Bi$_{2-x}$Sn$_x$Te$_2$Se$_{0.995}$ led to the simultaneous presence of $p$- and $n$-type carriers in the samples, resulting in the compensation effect, and $n-p$ crossover was observed while increasing $x$. 
For small Sn concentrations ($x\leq 0.005$), the samples were $n$-type, with relatively large thermopower and power factor. At around $x = 0.01$, the system was compensated (small Hall carrier concentration due to the opposite $n$ and $p$ components, the smallest $S$ and PF in the series). For higher $x$, holes delivered by Sn started to dominate, and the sample with $x = 0.04$ was $p$-type, but the absolute value of $S$ and PF were smaller, comparing to the $n$-type samples.
Authors summarized that work by saying that no evidence for the presence of Sn resonant level was found, and that Sn level might lie at lower energies, thus heavier $p$-type doping would have been required to observe the RL effects. 
In two other works, however~\cite{bi2te2se-sn-prb,bi2te2se-sn-cava1}, the presence of the impurity band, related to the Sn resonant level, was proposed, basing on other physical properties measurements, and supercell electronic structure computations.

\subsection{Se/Te vacancies}

\begin{figure}[t]
\includegraphics[width=0.50\textwidth]{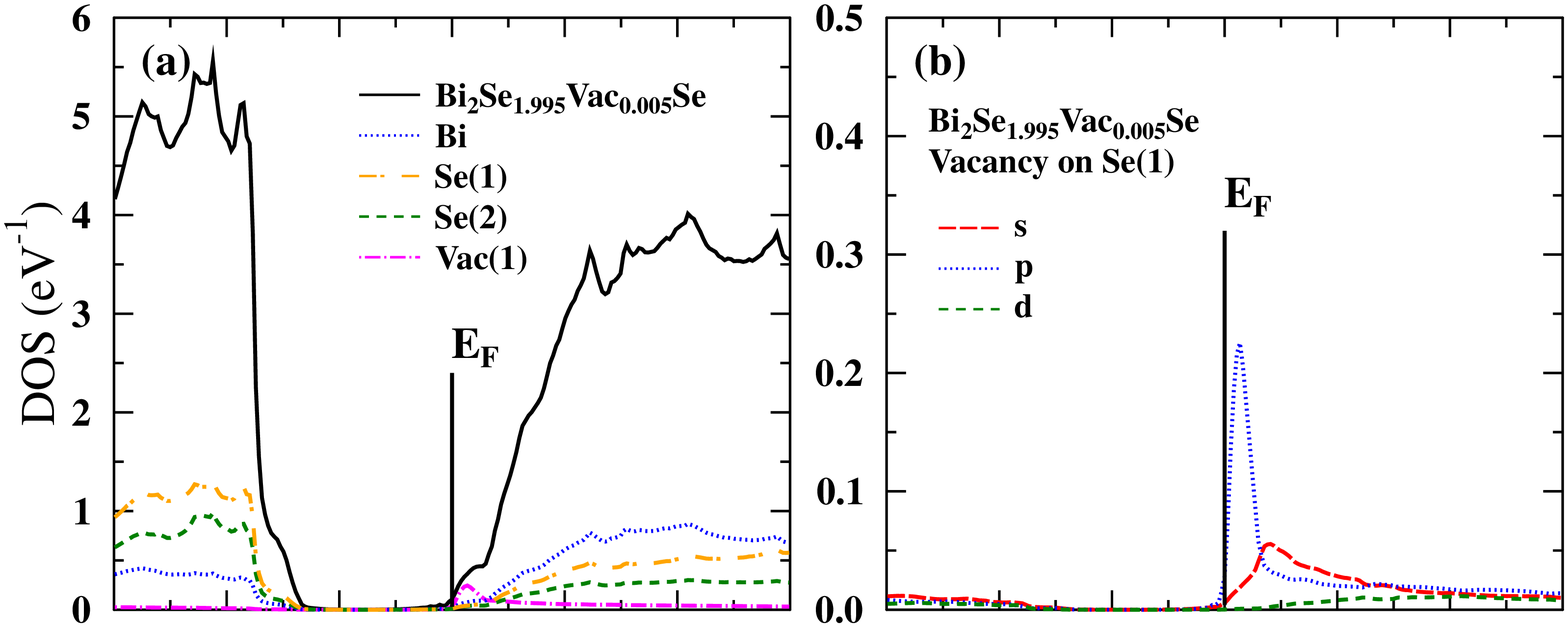}\\
\includegraphics[width=0.50\textwidth]{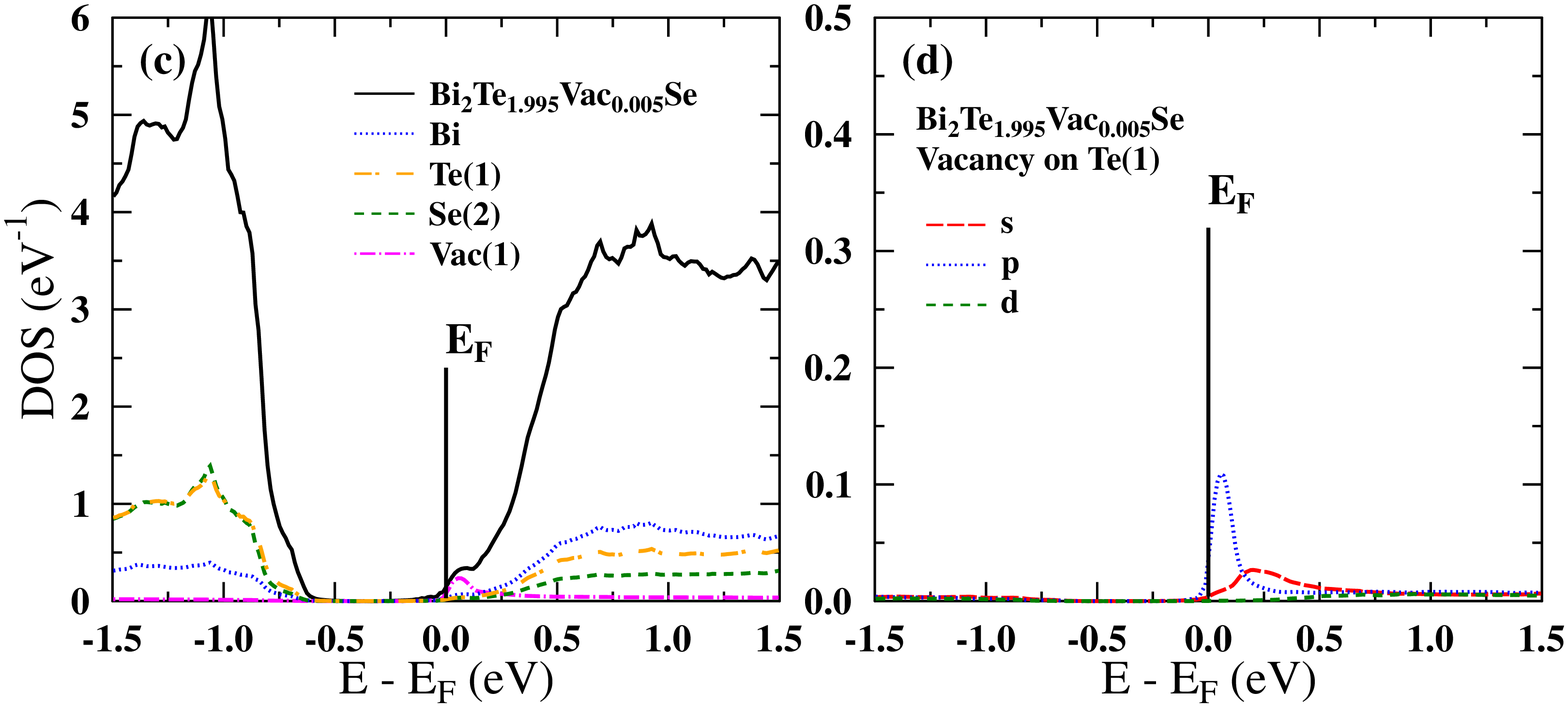}
\caption{Vacanies in the 'outer' (van der Waals) layer: (a) Total DOS of Bi$_{2}$Se$_{2.995}$ containing 0.5\% of vacancies on Se(1) atoms, located at (2a) crystal sites.  (b) partial DOS at vacancy, decomposed over the angular momentum; (c)-(d) same as (a)-(b) but for the Bi$_{2}$Te$_{1.995}$Se case, containing 0.5\% of vacancies on Te(1) atoms.}
\label{fig:vac2}
\end{figure}

\begin{figure}[htb]
\includegraphics[width=0.50\textwidth]{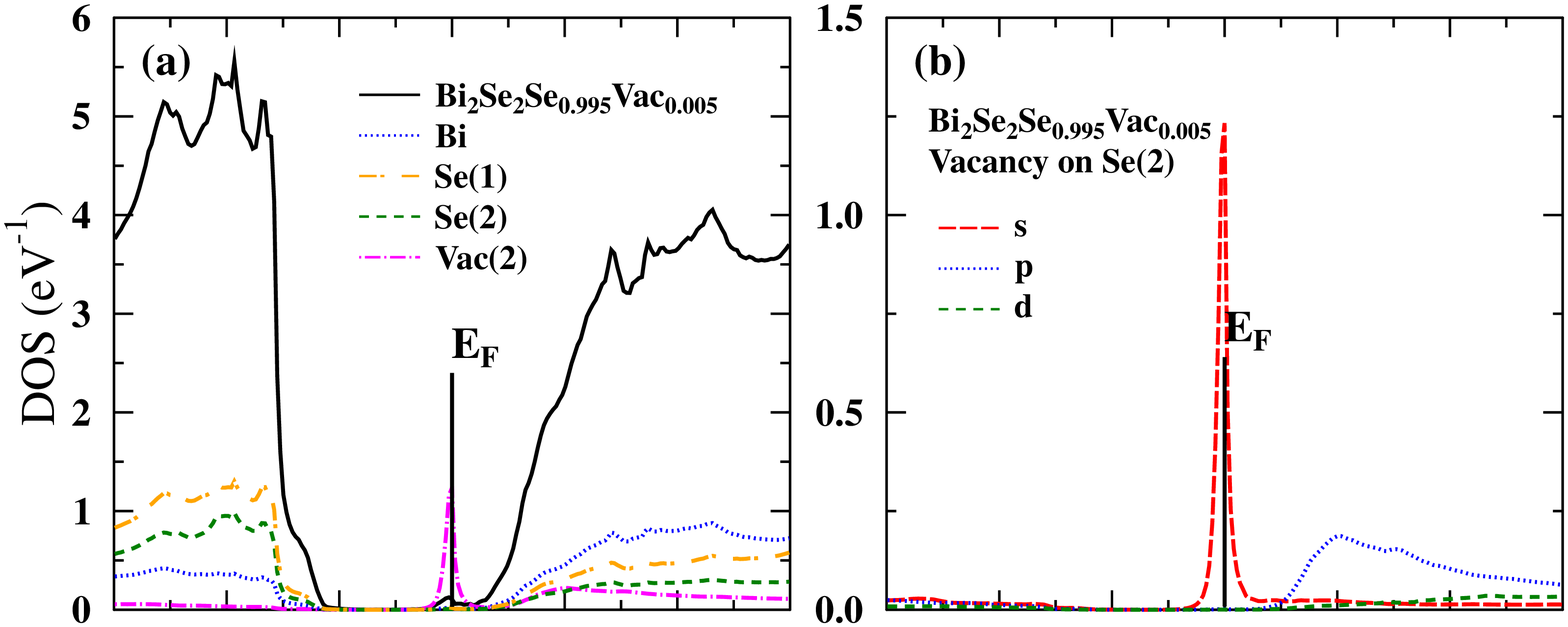}\\
\includegraphics[width=0.50\textwidth]{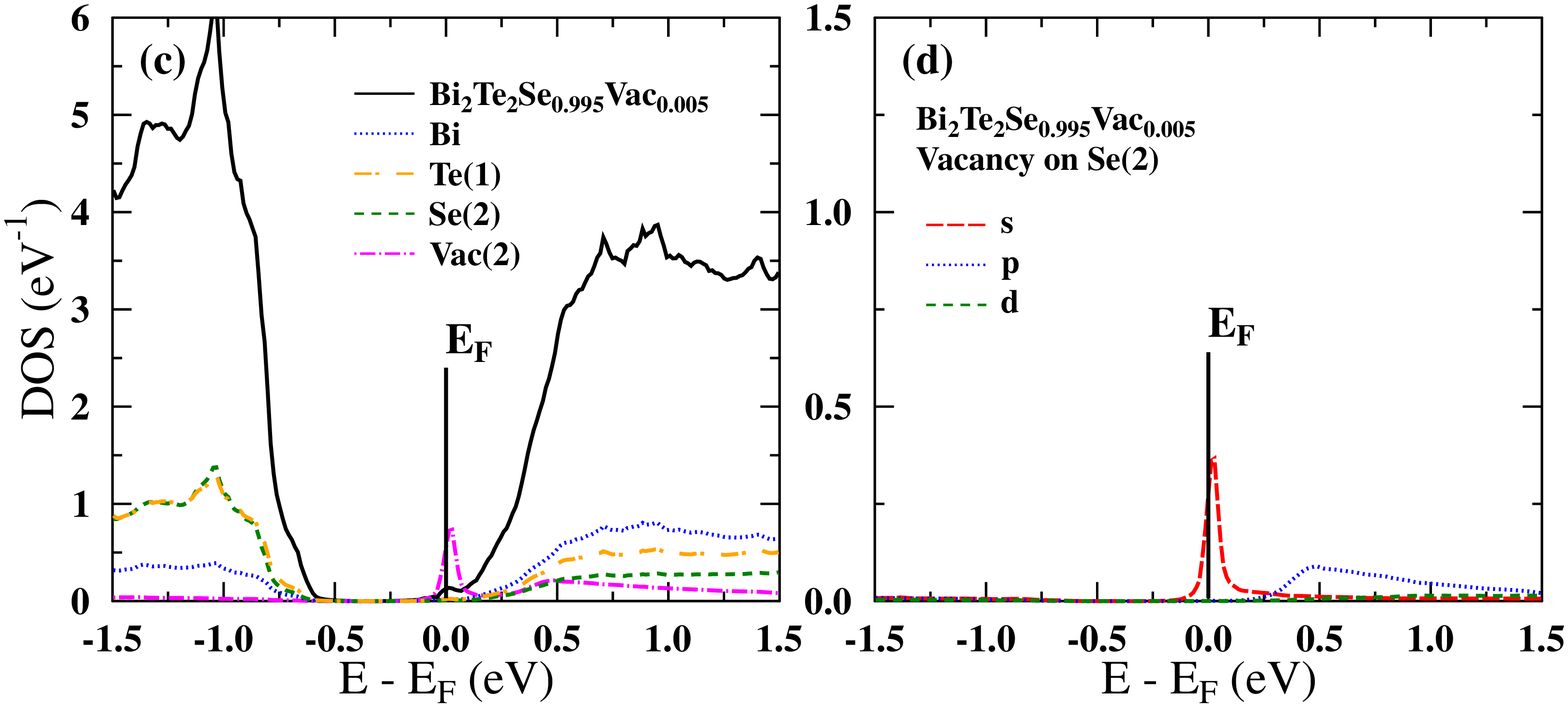}
\caption{Same as Fig.~\ref{fig:vac2} but for vacancy located in the inner layer: (a) Total DOS of Bi$_{2}$Se$_{2.995}$ containing 0.5\% of vacancies on Se(2) atoms, located at (1a) crystal sites; (b) partial DOS at vacancy, decomposed over the angular momentum; (c)-(d) same as (a)-(b) but for the Bi$_{2}$Te$_2$Se$_{0.995}$ case, containing 0.5\% of vacancies on Se(2) atoms.}
\label{fig:vac1}
\end{figure}

To verify whether formation of the resonant level on Sn may be altered by the presence of vacancies, electronic structure calculations for Bi$_2$Se$_3$ and Bi$_2$Te$_2$Se containing simultaneously vacancies and impurity atoms, have to be performed. 
We start our analysis considering sole vacancies on chalcogen (Se, Te) atoms, followed by more complex cases of the doped and doubly-doped systems containing vacancies.
In general, there are two possibilities of location of vanancies on the chalcogen atoms in the rhombohedral tetradymite. The first one is the vacancy in the 'outer' (van der Waals) atomic layer, formed by Se(1) in Bi$_2$Se$_3$ and Te(1) in Bi$_2$Te$_2$Se, which we label in this work as Vac(1).
The second one is the vacancy in the 'inner' atomic layer, formed by Se(2) in Bi$_2$Se$_3$ and Bi$_2$Te$_2$Se, which we label in this work as Vac(2).
Intuitively, the weakly-bounded outer layer should be more predisposed to contain vacancies.
Theoretical studies on various types of defects in the tetradymites were recently reported (see e.g. Refs.~\cite{def3,def1,def2}), and they generally confirm this prediction, i.e. more probable (at least under Bi-rich conditions) are vacancies on Se(1) in Bi$_2$Se$_3$ and on Te(1) in Bi$_2$Te$_2$Se, rather than on Se(2).
It is worth noting, that the second prediction of the more probable vacancies formation on Te, rather than on Se in Bi$_2$Te$_2$Se, disagree with the frequently discussed model~\cite{cava-tetra_cryst,bi2te2se-sn-fuccillo}.

To make our results more general and independent of assumed vacancy location, in both, Bi$_2$Se$_3$ and Bi$_2$Te$_2$Se, two possibilities of vacancies location were considered.
As will be shown next, the most important observation is that the actual location of the vacancy is not critically important, especially if the material is desired to be $p$-type (and vacancies have to be counter-doped). The cumulative vacancy concentration per formula unit is the key parameter controlling the type of conductivity in the system.

Electronic structures of compounds with vacancies were calculated, using the same KKR-CPA method, as in previous paragraph. Technically, vacancy was simulated as an ''empty sphere'', i.e. an ''atom'' with $Z = 0$, placed on the selected site, with simulated concentration, and electronic structure was then calculated self-consistently, using the coherent potential approximation.

First, we will describe the more intuitive, outer-layer vacancies case, Vac(1). Figures~\ref{fig:vac2}(a)-(b) show DOS of Bi$_2$Se$_3$ with 0.5\% of vacancies on Se(1) atoms, whereas Figures~\ref{fig:vac2}(c)-(d) presents the case of Bi$_2$Te$_2$Se with 0.5\% of vacancies on Te(1). First of all, calculations correctly predict the $n$-type conductivity of the materials, with $E_F$ located in the conduction band, as if the vacancies were electron donors.
However, rather striking feature is also observed in the Figures. These are small bumps of DOS created by the vacancies, seen for both cases. 
The partial DOS peaks on vacancies are several times smaler, than RL peaks on Sn atoms in these structures, but still, vacancies show a non-rigid band like behavior, even forming resonant-like state, which may modify the electronic structure close to the conduction band edge.

Similar DOS peaks are observed for the second possibility of the vacancy location, Vac(2), i.e. in the 'inner' Se(2) atomic layer, as presented in Fig.~\ref{fig:vac1}. For this case, DOS peaks on vacancies are even larger and more narrow, which might indicate that these crystal sites are not preferred for the vacancy location, in agreement with the results of Refs.~\cite{def3,def1,def2}.
Another difference between Vac(1) and Vac(2) is visible, when angular momentum decomposition of densities of states is compared in Figures~\ref{fig:vac2}(b),(d) and Figures~\ref{fig:vac1}(b),(d).The DOS peak at Vac(1), i.e. in the outer layer, has mainly p-like orbital momentum character, in contrast to the s-like states involved in the DOS peak at Vac(2). 
Nevertheless, the global effect of the presence of vacancies on the electronic structure is similar: both cases lead to $n$-type conductivity and vacancies behave as charged donors (more precisely, two electron donors, as will be clarified in the next paragraph). For present work and behavior of materials, which are designed to contain $p$-type RL impurities, like Sn, compensation of these donors is required, and the difference between Vac(1) and Vac(2) is not important, as long as they do not alter the RL formation, which we will address next.
The key parameter, controlling the Fermi level position, is the total concentration of the chalcogen vacancies, and the same number of additional acceptor atoms is required to compensate them, regardless of the site they occupy.

It is worth noting, that this rather unexpected, resonant-like behavior of the vacancies, remains in agreement with the experimental studies, since the presence of resonant electronic states on Se vacancies in Bi$_2$Se$_3$ was earlier found experimentally, using the scanning tunneling microscopy (STM)~\cite{bi2se3-vac-res}.

Our results show, that vacancies may have stronger, than expected, influence on the electronic structure of $n$-type Bi$_2$Se$_3$ and Bi$_2$Te$_2$Se, which may go beyond the simple rigid shift of the Fermi level into the conduction band. 
At present, it is however difficult to judge, whether vacancies have any considerable, positive or negative effect on the thermoelectric properties of these materials, since there are no vacancy-free samples to compare with. 
For sure, mobility of electrons is expected to slightly decrease, when vacancies are present in a sample, due to scattering on defects, which would be degrading to thermoelectric performance. 
If any resonant-like, positive effect on the thermopower would be associated with the local increase in DOS, 
we may expect that it should be relatively weaker, than due to the RL from Sn atoms, since the resulting DOS, associated with the vacancy, is considerably smaller. Thus, the balance between the drop in the mobility against the (potential) gain in the thermopower may result in a little effect on the power factor. 
But one cannot exclude that the relatively large values of the thermopower in $n$-type Bi$_2$Te$_2$Se containing vacancies are correlated with the formation of the 'hump' in DOS due to the presence of vacancies. Further experimental efforts on synthesis of vacancy-free samples would be required to draw any further conclusions.

\subsection{Se/Te vacancies plus regular dopants}

\begin{figure}[b]
\includegraphics[width=0.50\textwidth]{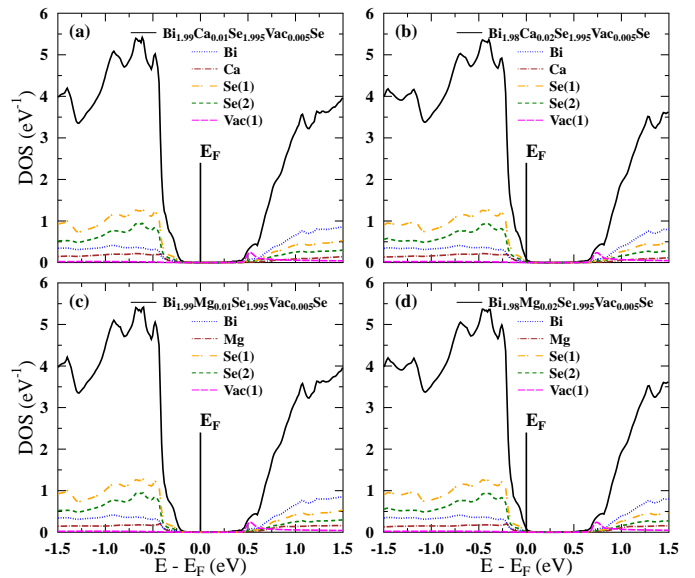}
  \caption{DOS of Ca (a-b) and Mg (c-d) doped Bi$_2$Se$_{2.995}$ for the case, where vacancy is in the outer Se(1) layer. At panels (a)/(c) 1\% of Ca/Mg compensates the donor behavior of 0.5\% of vacancies, for larger Ca/Mg concentrations material becomes $p$-type, as shown in (b)/(d).}
  \label{fig:S2}
\end{figure}

\begin{figure}[t]
\includegraphics[width=0.50\textwidth]{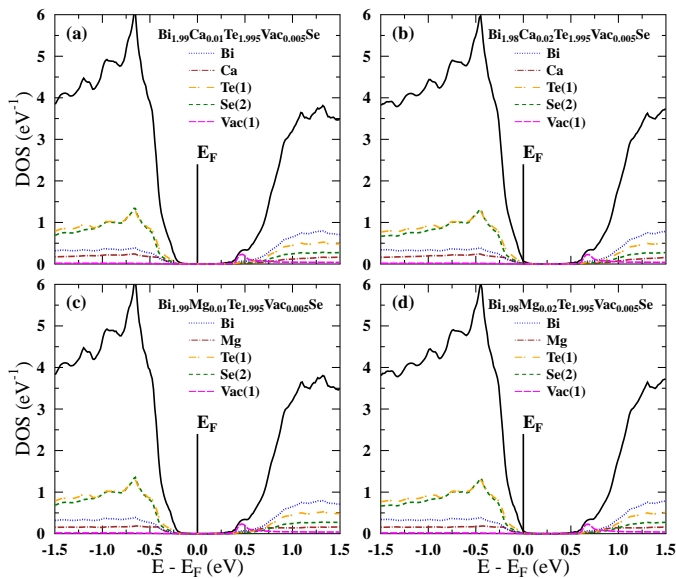}
  \caption{DOS of Ca (a-b) and Mg (c-d) doped Bi$_2$Te$_{1.995}$Se ('outer' layer Vac(1) case). At panels (a)/(c) 1\% of Ca/Mg compensates the donor behavior of 0.5\% of vacancies, for larger Ca/Mg concentrations material becomes $p$-type, as shown in (b)/(d).}
  \label{fig:S8}
\end{figure}

The successful way of compensating vacancies, tuning of the Fermi level and making Bi$_2$Se$_3$ $p$-type was recently established to be by Ca~\cite{cava-bi2se3-prb09}, Mg~\cite{mg-bi2se3}, or Mn~\cite{mn-bi2se3,bi2se3-gao} doping on Bi site. Since Mn, with its open 3d shell and magnetic properties would require a separate discussion, in this work the two alkaline-earth elements are only studied.
Our calculations results for Ca and Mg doped systems containing vacancies in the outer layer (Vac(1)), are presented in Fig.~\ref{fig:S2} for Bi$_2$Se$_3$ and  Fig.~\ref{fig:S8} for Bi$_2$Te$_2$Se. The same set of figures for the vacancies in the inner layer (Vac(2) case) are presented in the Appendix, Fig.~\ref{fig:vaccamgse}
and Fig.~\ref{fig:vaccamgtese}, respectively.
In all the cases, Mg and Ca behave as regular acceptors, rigidly moving the Fermi level towards the valence band, without noticeable changes in DOS of the doped materials. Both of the alkaline earth elements behave as single electron acceptors, delivering one hole per substituted atom. When the concentration of the dopant is equal to two times the concentration of vacancies (see, Figs.~\ref{fig:S2}-\ref{fig:S8} and Figs.~\ref{fig:vaccamgse}-\ref{fig:vaccamgtese}) the material is compensated and $E_F$ is placed in the middle of the band gap. For larger concentrations of the acceptors, effectively $p$-type material is obtained, in agreement with experiment, and for both Vac(1) and Vac(2) locations.

\subsection{Se/Te vacancies plus Sn}

Now the question arises, how the resonant state on Sn behave, when vacancies are present in the material.
As we mentioned before, the thermoelectric properties of Bi$_2$Te$_2$Se:Sn were systematically investigated experimentally, and no substantial effect on the thermopower, due to the presence of Sn, was found~\cite{bi2te2se-sn-fuccillo}.
Now we will try to reproduce that experiment and see how the electronic structure changes, when Sn is added to the vacancy-containing compound. 
In Ref.~\cite{bi2te2se-sn-fuccillo} authors assumed that the $n$-type behavior of initial Bi$_2$Te$_2$Se material is due to the presence of 0.5\% vacancies in the inner chalcogen layer (i.e. on Se(2) atoms, our Vac(2) case). We show below, that the material's behavior does not depend on this assumption, and similar changes in transport properties of Bi$_2$Te$_2$Se could be observed if the vacancies were located in the outer layer of Te atoms.

\begin{figure}[t]
\includegraphics[width=0.50\textwidth]{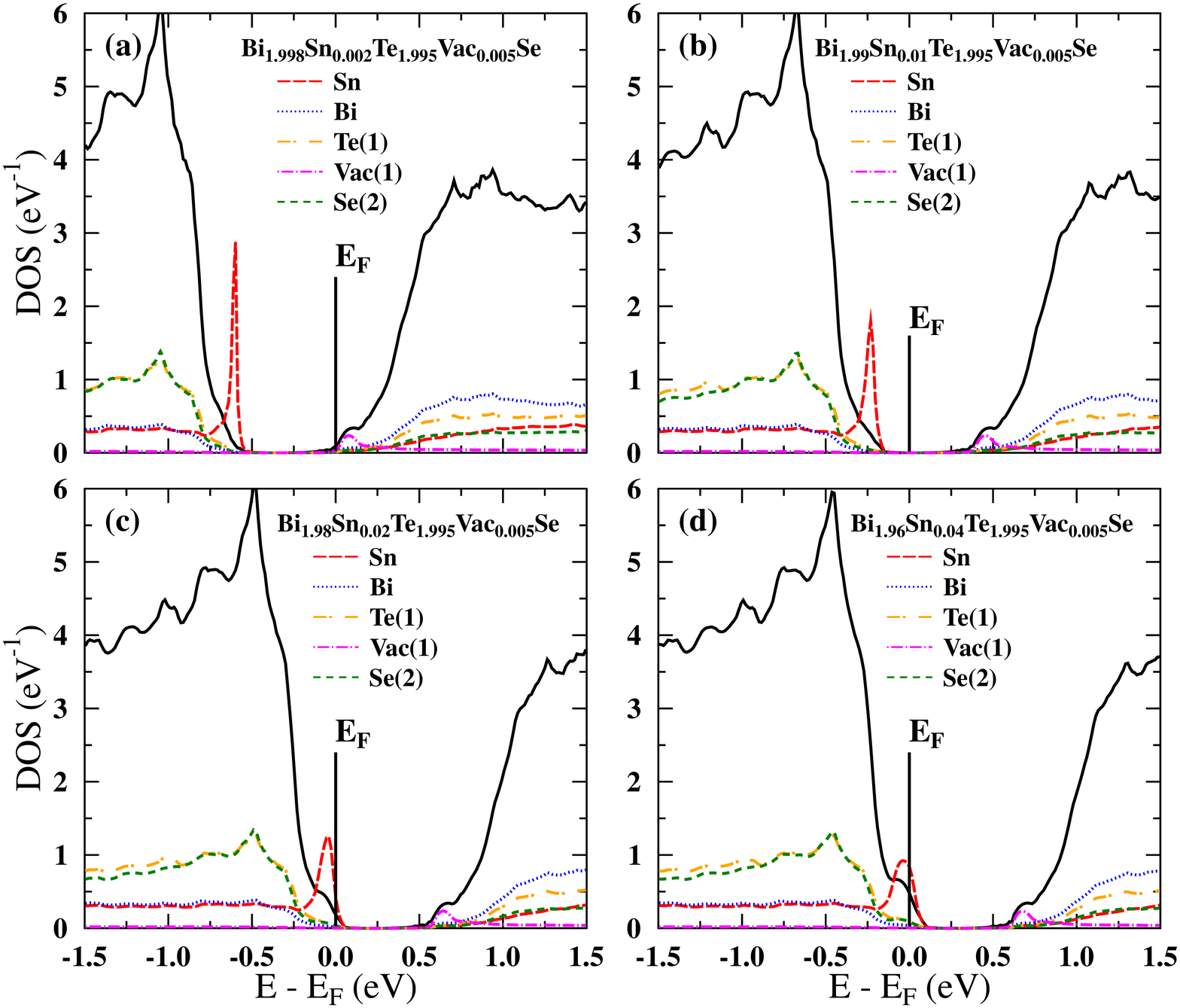}  
  \caption{Evolution of DOS and Fermi level position in Sn doped Bi$_2$Te$_{1.995}$Se. When Sn concentration is larger than two times the vacancy concentration, $E_F$ starts to penetrate the RL DOS peak (panel c), however for large concentrations (panel d) RL peak becomes rather broad.}
  \label{fig:snvactese}
\end{figure}

\begin{figure}[t]
\includegraphics[width=0.25\textwidth]{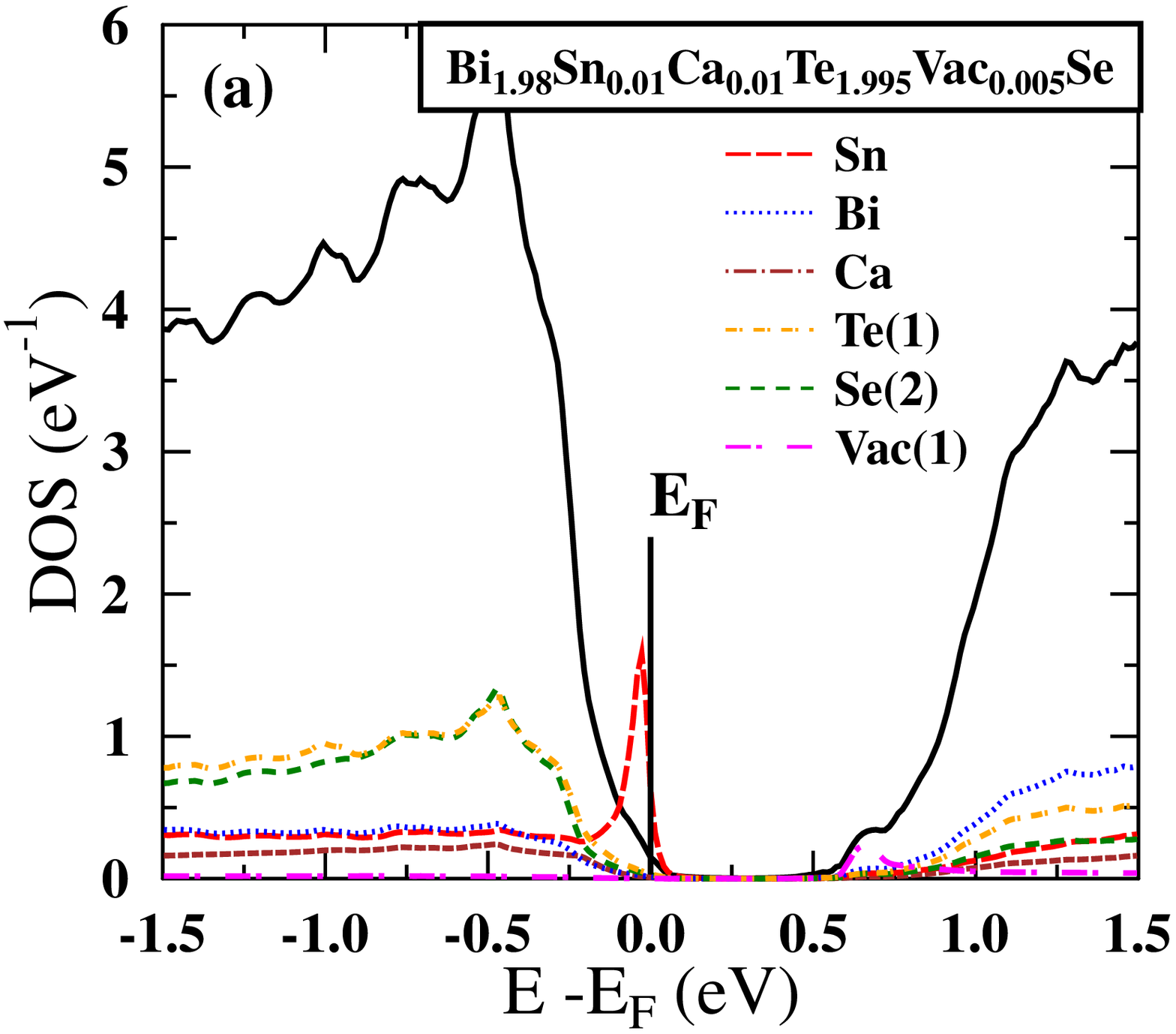}\includegraphics[width=0.24\textwidth]{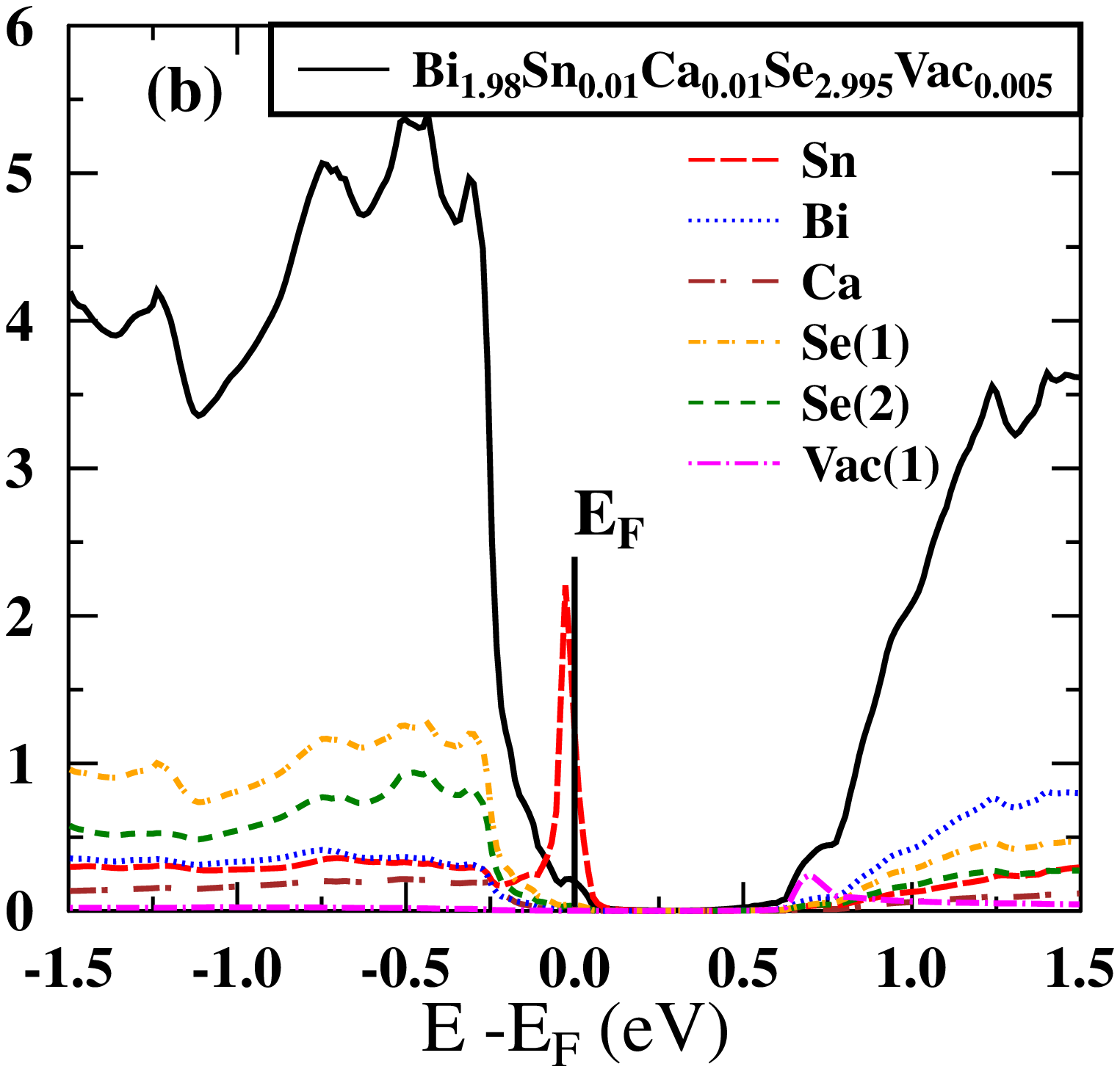}
  \caption{DOS of the doubly, Sn- and Ca-doped (a) Bi$_2$Te$_2$Se, and (b) Bi$_2$Se$_{3}$, both containing 0.5\% of the outer-layer vacancies Vac(1). Our results show that the double doping, using a rigid-band-like acceptor (here Ca), should be an effective way of tuning the Fermi level position, without broadening the Sn RL too much.}
  \label{fig:S5}
\end{figure}

As for previous analysis, we start with considering the more intuitive, outer layer vacancies (Vac(1)). Figure~\ref{fig:snvactese} shows the evolution of the density of states of Bi$_{2-x}$Sn$_x$Te$_{1.995}$Se, i.e. for the fixed 0.5\% Te(1) vacancy concentration. In panel (a), for $x=0.002$ of Sn, the system is $n$-type, and large Sn resonant peak is present on the valence side of the band gap. For $x=0.01$ (Fig.~\ref{fig:snvactese}(b)) the holes delivered by Sn lead to the compensation of the donor vacancy effect, and $E_F$, as in the Ca and Mg cases, falls into the gap. 
Further increase of the tin concentration, $x=0.02$ (panel (c)), and $x=0.04$ (panel (d)) places $E_F$ inside the valence band. 
The Sn concentration, where $n-p$ crossover takes place, remains in a very good agreement with the experimental studies~\cite{bi2te2se-sn-fuccillo}.
Exactly the same behavior is observed, when the vacancy occupies the inner (Se) atomic layer (see, Appendix, Fig.~\ref{fig:S7}).

From our calculations we see, that the formation of the RL on Sn is not disturbed by the presence of vacancies, and $E_F$ should penetrate the DOS region, where the RL from Sn is present also for the 4\% Sn doped sample, studied in Ref.~\cite{bi2te2se-sn-fuccillo}. 
The problem that the effect of RL was not observed in the thermopower measurements may be related to the large concentration of Sn, used in that experiment. Earlier studies on Bi$_2$Te$_3$:Sn, as we mentioned in Sec.~\ref{sec:sn_te}, showed, that Sn concentration has to precisely tuned, to observe the increase of thermopower. 
In Bi$_2$Te$_3$:Sn it was done for $1\%\leq x \leq 2\%$, and for the larger Sn concentration of $x = 5\%$, the increase in thermopower was no longer evident. 
In Ref.~\cite{bi2te2se-sn-fuccillo} the authors had to heavily dope the samples with tin to overcome the $n$-type behavior driven by defects, and succeeded in obtaining a $p$-type sample for $x = 4\%$. It is possible, that such concentration may be too large to observe the positive effect of Sn on the thermopower, similarly to the 5\% Sn doped Bi$_2$Te$_3$ -- for too large amount of dopants, the RL may become too broad and Sn may loose its resonant character, starting to behave as a typical acceptor.

\begin{figure}[htb]
\includegraphics[width=0.50\textwidth]{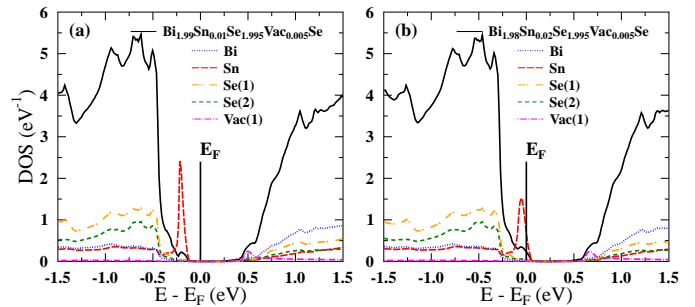}
  \caption{Evolution of DOS and Fermi level position in Sn doped Bi$_2$Se$_{2.995}$ for the case, where vacancy is on Se(1) atoms ('outer layer' Vac(1)). When Sn concentration is two times the vacancy concentration (panel (a)), $E_F$ material is compensated. For larger concentrations (panel (b)) $E_F$ is in the valence band, but RL peak broadens.}
  \label{fig:S3}
\end{figure}

We are now in position to suggest, at least in principle, a simple way of resolving this problem. 
Instead of heavily dope the sample with Sn to reach the valence band, one can use the regular acceptors, like Ca or Mg, to compensate the $n$-type defects and tune the Fermi level position, keeping the Sn concentration constant at lower level, lets say between $x = 0.5\% - 2\%$. Measurements on such a series of the doubly-doped samples should clarify whether the Sn resonant level in Bi$_2$Te$_2$Se can be reached and whether it has a positive influence on the thermoelectric properties of the material. Figure~\ref{fig:S5}(a) shows DOS of such a doubly doped system, containing 0.5\% Te vacancies: Bi$_{1.98}$Sn$_{0.01}$Ca$_{0.01}$Te$_{1.995}$Se. For the 1\% Sn concentration and without Ca, the material would be compensated, as in Fig.~\ref{fig:snvactese}(b). Addition of 1\% of calcium shifts the  Fermi level into the valence band, and Sn keeps its substantial partial DOS peak, which is not smeared much as would be observed for the larger Sn concentration.
Thus, double-doping seems to be an efficient way of controlling the Fermi level position in the presence of vacancies and resonant level. 
Moreover, the same conclusion holds for the second possibility of the vacancy location -- see, Appendix, Fig.~\ref{fig:sncavacse}(a) for similar DOS picture for the Vac(2) case.

As far as Bi$_2$Se$_3$ is concerned, we are not aware of any published experimental works on its thermoelectric properties upon Sn doping. Our computations show, that theoretically, conclusions drawn above for Bi$_2$Te$_2$Se are also valid for vacancy-containing Bi$_2$Se$_3$, and formation of the RL on Sn in DFT computations is independent of the presence of vacancy. 
The evolution of DOS and Fermi level with increasing Sn concentration in Bi$_{2-x}$Sn$_x$Se$_{2.995}$ was found to be similar to the Sn-doped Bi$_2$Te$_{1.995}$Se case (Fig.~\ref{fig:snvactese}), thus in Fig.~\ref{fig:S3} we show only DOS for 1\% and 2\% Sn concentrations, where the material is compensated (panel(a)), and where $n$-$p$ crossover takes place (panel (b)).
Again, as for Bi$_2$Te$_2$Se, we show in Fig.~\ref{fig:S5}(b), that the double-doping strategy may be used to effectively control the Fermi energy, keeping Sn concentration at the level of 1-2\%, where positive effects of RL on the thermopower may be expected. 
These aforementioned observations do not change when vacancy is located in the inner layer (Se(2) atoms, Vac(2) case), which is presented in Appendix, Fig.~\ref{fig:snvacse} and Fig.~\ref{fig:sncavacse}(b).

\subsection{Al and Ga - possible new resonant impurities}

\begin{figure}[b]
\includegraphics[width=0.50\textwidth]{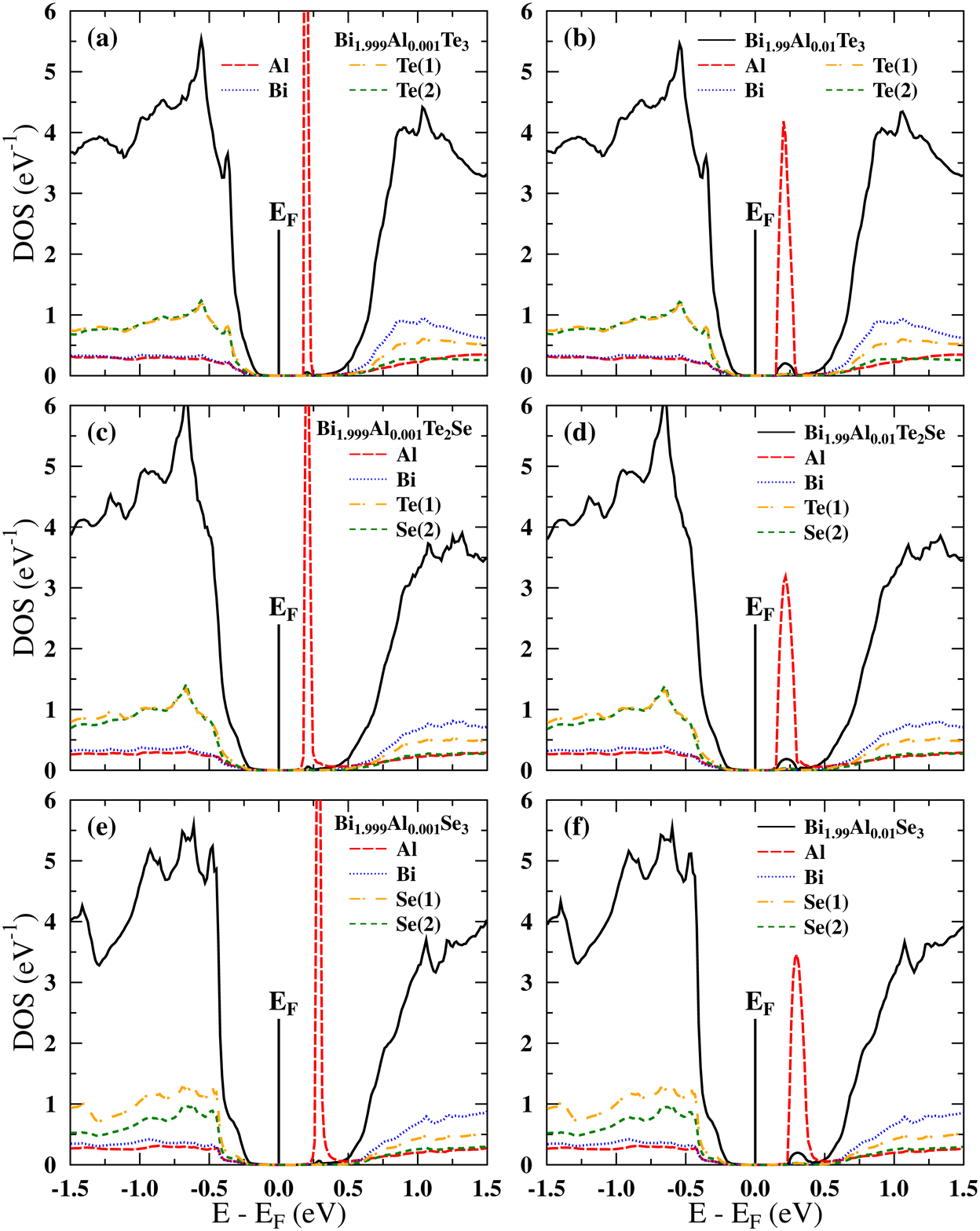}
  \caption{DOS of Al doped tetradymites: (a-b) Bi$_2$Te$_3$; (c-d) Bi$_2$Te$_2$Se; (e-f) Bi$_2$Se$_3$. Strongly peaked DOS shows the possibility of RL formation, but $E_F$ position is not sensitive to the concentration of Al, thus Al may be an electrically inactive impurity.}
  \label{fig:altetra}
\end{figure}

\begin{figure}[t]
  \includegraphics[width=0.50\textwidth]{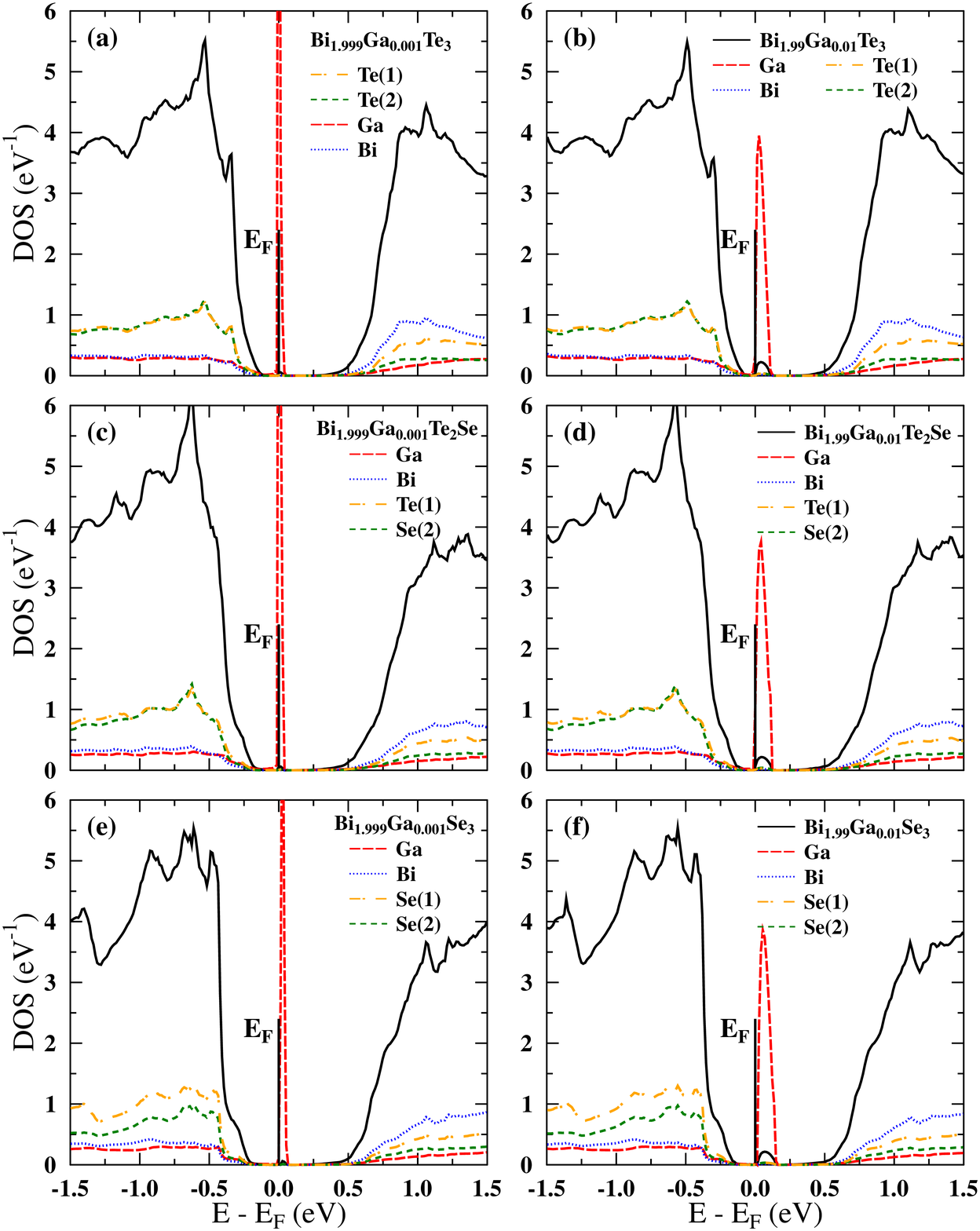}
  \caption{DOS of Ga doped tetradymites: (a-b) Bi$_2$Te$_3$; (c-d) Bi$_2$Te$_2$Se; (e-f) Bi$_2$Se$_3$. Similarly to Al, strongly peaked DOS shows the possibility of RL formation, and position of $E_F$ does not change with Ga concentration. If position of Ga level is correctly predicted by LDA calculations, Ga is more pronounced to form a trap state in the gap, than Al.}
  \label{fig:gatetra}
\end{figure}

In the last section of this work we would like to suggest two new possible resonant impurities in the tetradymite series of materials, namely Al and Ga, to stimulate further experimental works on them. As far as the literature reports are concerned, two works~\cite{ga-bisbte3-1,ga-bisbte3-2} reported anomalous increase of the thermopower and $ZT$ in Ga doped $n$-type (Bi$_{0.5}$Sb$_{0.5}$)$_2$Te$_3$, alloy, which may be due to the RL formation. 
Results of the calculations for the series of Al and Ga doped materials are presented in Fig.~\ref{fig:altetra} and Fig.~\ref{fig:gatetra}.

For both, Al (Fig.~\ref{fig:altetra}) and Ga (Fig.~\ref{fig:gatetra}) impurities, and all of the host materials, we observe the resonant peaks in the partial DOS of the impurity atoms. As in Sn case, DOS peaks are $s$-like, 
however several important differences are noticed. 
For the Al case, resonant level is located at the edge of the conduction band (CB), and moves further towards CB when electronegativity of the neighboring atoms increase (from Te to Se). Evolution of the DOS with increasing Al concentration shows, that actually Al is not changing the Fermi level position, which is located in the gap. Thus, Al seems to be an isoelectronic impurity, which is understood as both, Al and Bi, are trivalent. Because of this, double-doping with a second donor may be required to tune the Fermi level position and scan the resonant peak.

For Ga doped materials (Fig.~\ref{fig:gatetra}), the position of Ga level is also in the gap, but closer to the valence band edge, if compared to Al. Similarly to that case, we do not observe any shift in the Fermi level position, when concentration of the impurity is changed: $E_F$ is rather pinned to the RL peak in DOS within the gap. 
Also here, double doping with another donor will be needed to properly place $E_F$ and investigate the Ga level.
Partial substitution of Bi atoms with Sb may also affect these RL positions, since it will slightly modifie the electronegativity of the cation site, as well as create chemical pressure. Both of these factors may influence the RL position.

The fact, that Al and Ga impurities are isoelectronic with Bi in tetradymites, may have an additional, positive effect on the thermoelectric performance. For such impurities we may expect that presence of Al or Ga will not induce any ionized impurity scattering, which would reduce the mobility of carriers in the system, but may additionally increase the thermopower via neutral impurity scattering, similarly as was observed in the In and Ga doped Bi~\cite{Jin2015}.

For these two dopants, we cannot directly compare our predictions with the experiment, since the only measurements were reported for (Bi-Sb)$_2$Te$_3$ alloy, which we do not model here, but possibility of the formation of the resonant level by Ga is strongly supported by the presented theoretical calculations.  
Additionally, we have to note, that our calculations are done using several approximations, including LDA, which can be not very accurate in predicting the exact position of the RL, especially for the case when it is above $E_F$. Here, the same problem occurs as for the band gap value, and the mutual relation between the RL position and the conduction band edge is influenced by inaccuracy of the gap. Thus, experimental verification of our predictions is highly desired.

\section{Summary and conclusions}

Results of the first principles calculations of the electronic structure of the tetradymites, Bi$_2$Te$_3$, Bi$_2$Te$_2$Se and Bi$_2$Se$_3$, containing various impurities and vacancies were presented. We have found, that Sn should be a resonant impurity in all aforementioned materials, which remains in agreement with the experimental findings for Bi$_2$Te$_3$. Studies on Se-containing materials, that is Bi$_2$Te$_2$Se and Bi$_2$Se$_3$, have shown that vacancies, which are very likely to be present in real samples, behave as charge donors and also create small, resonant-like peaks in the densities of states near the bottom of the conduction band. The angular momentum character of DOS peaks on vacancies depend on which chalcogen atom site they are located on, however overall characteristics of the system, like the sign of majority carriers, depend only on the total vacancy concentration.
Doping the Bi site with Ca or Mg, which was earlier experimentally shown to result in $p$-type materials,
is theoretically confirmed as a rigid-band-like way of controlling the position of the Fermi level and turning the defected, $n$-type Bi$_2$Te$_2$Se and Bi$_2$Se$_3$, into $p$-type, with no major changes in the valence band DOS.
For Sn doped Bi$_2$Te$_2$Se and Bi$_2$Se$_{3}$ containing vacancies, we have shown, that co-doping of the material with Ca or Mg should be a better way of compensating the donor behavior of Te/Se vacancies, to reach the Sn resonant level, than by heavy doping with Sn itself. This should help to avoid smearing of the resonant level, which is observed for the larger Sn concentrations, and for which the resonant character of Sn may be lost. 
Finally, we have suggested the two new, potentially interesting impurities Al and Ga, which are predicted to form resonant levels in the Bi-based tetradymites.

\section*{Acknowledgements}
This work was partially supported by the Polish National Science Center (NCN) (project no. DEC-2011/02/A/ST3/00124) and the Polish Ministry of Science and Higher Education. 

\newpage

\section*{Appendix \\ Second choice of the location of chalcogen atom vacancy site in Bi$_2$Se$_3$ and Bi$_2$Te$_2$Se}

\begin{figure}[b]
\includegraphics[width=0.50\textwidth]{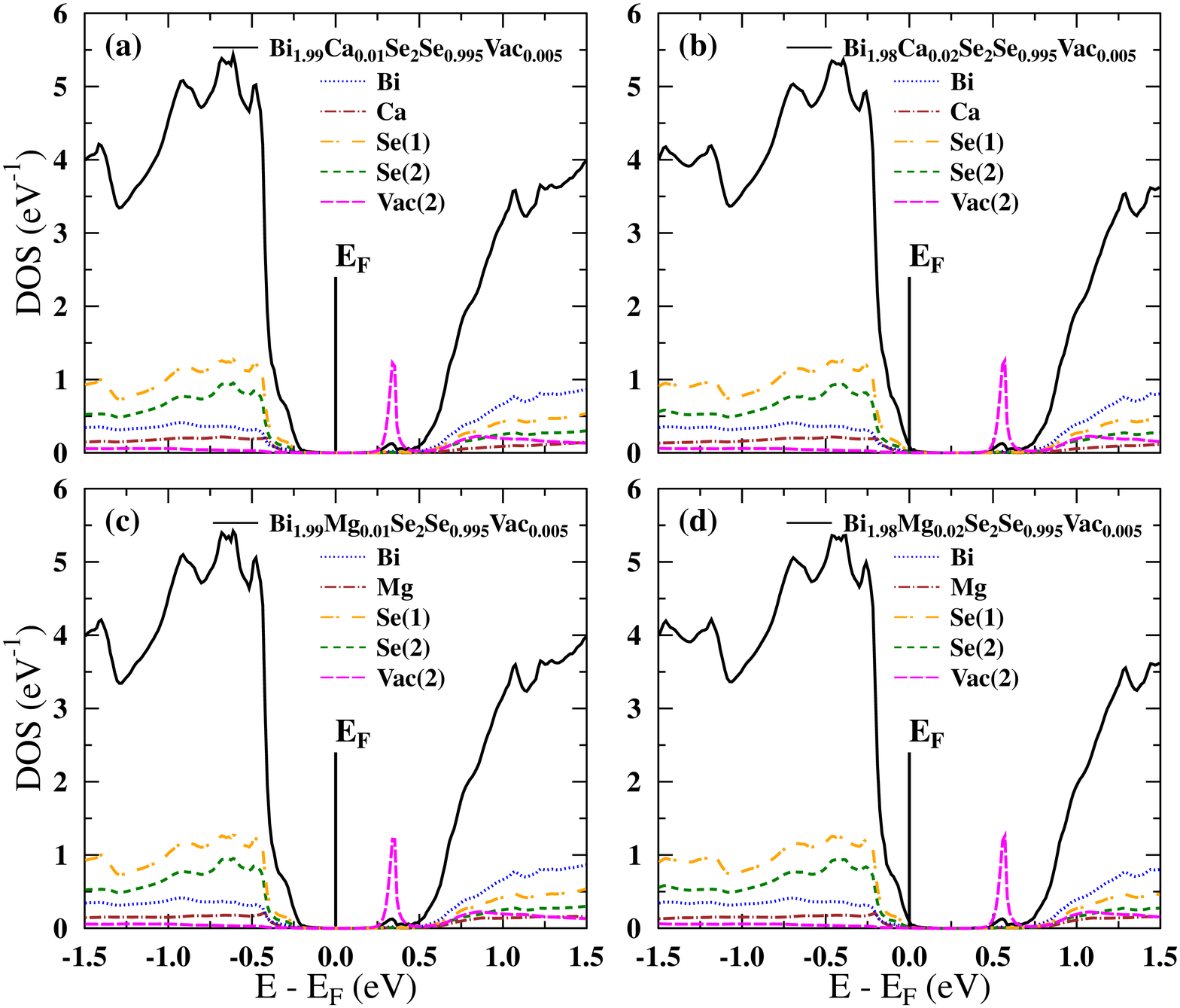}
\caption{'Inner layer' vacancy case for Ca (a-b) and Mg (c-d) doped Bi$_2$Se$_{2.995}$, i.e. containing 0.5\% of Se(2) vacancies. At panels (a)/(c) 1\% of Ca/Mg compensates the donor behavior of 0.5\% of vacancies, for larger Ca/Mg concentrations material becomes $p$-type, as shown in (b)/(d). $E_F$ behavior is exactly the same, as for the 'outer layer' Vac(1) case, presented in Fig.~\ref{fig:S2}}
\label{fig:vaccamgse}
\end{figure}
\begin{figure}[b]
\includegraphics[width=0.50\textwidth]{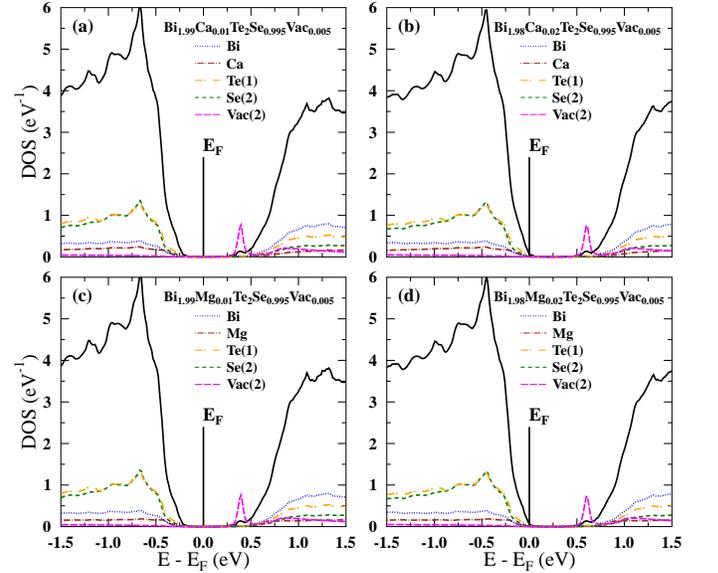}
  \caption{'Inner layer' vacancy case for Ca (a-b) and Mg (c-d) doped Bi$_2$Te$_2$Se$_{0.995}$, i.e. containing 0.5\% of Se(2) vacancies. 
  At panels (a)/(c) 1\% of Ca/Mg compensates the donor behavior of 0.5\% of vacancies, for larger Ca/Mg concentrations material becomes $p$-type, as shown in (b)/(d). Evolution of the electronic structure is similar to the 'outer layer' Vac(1) case, presented in Fig.~\ref{fig:S8}}
  \label{fig:vaccamgtese}
\end{figure}

In the Appendix we present the DOS figures, resulting from the electronic structure calculations for the doped Bi$_2$Se$_3$ and Bi$_2$Te$_2$Se containing vacancies in the 'inner' Se(2) chalcogen atomic layer (our Vac(2) case).
Comparison with the figures included in Section~\ref{sec:results} show that the main results of this work do not depend on the position of vacancy, i.e. we observe: (i) similar, two-electron-donor character of the vacancy, which forms small resonant-like peaks in DOS; (ii) formation of RL at Sn also in the presence of vacancies; (iii) possibility of the defect compensation upon doping with Mg and Ca, with the same acceptor concentration at the n-p crossover; (iv) successful tuning of the Fermi level position by double-doping of the moderately Sn-doped vacancy-containing materials, which helps to avoid the RL smearing effect, likely present for the heavily Sn-doped samples.
These results show, that the vacancy concentration, not the vacancy location, is the key parameter for the properties of Bi$_2$Te$_2$Se and Bi$_2$Se$_3$, if one wishes to prepare a $p$-type material.

\newpage

\begin{figure}[b]
\includegraphics[width=0.50\textwidth]{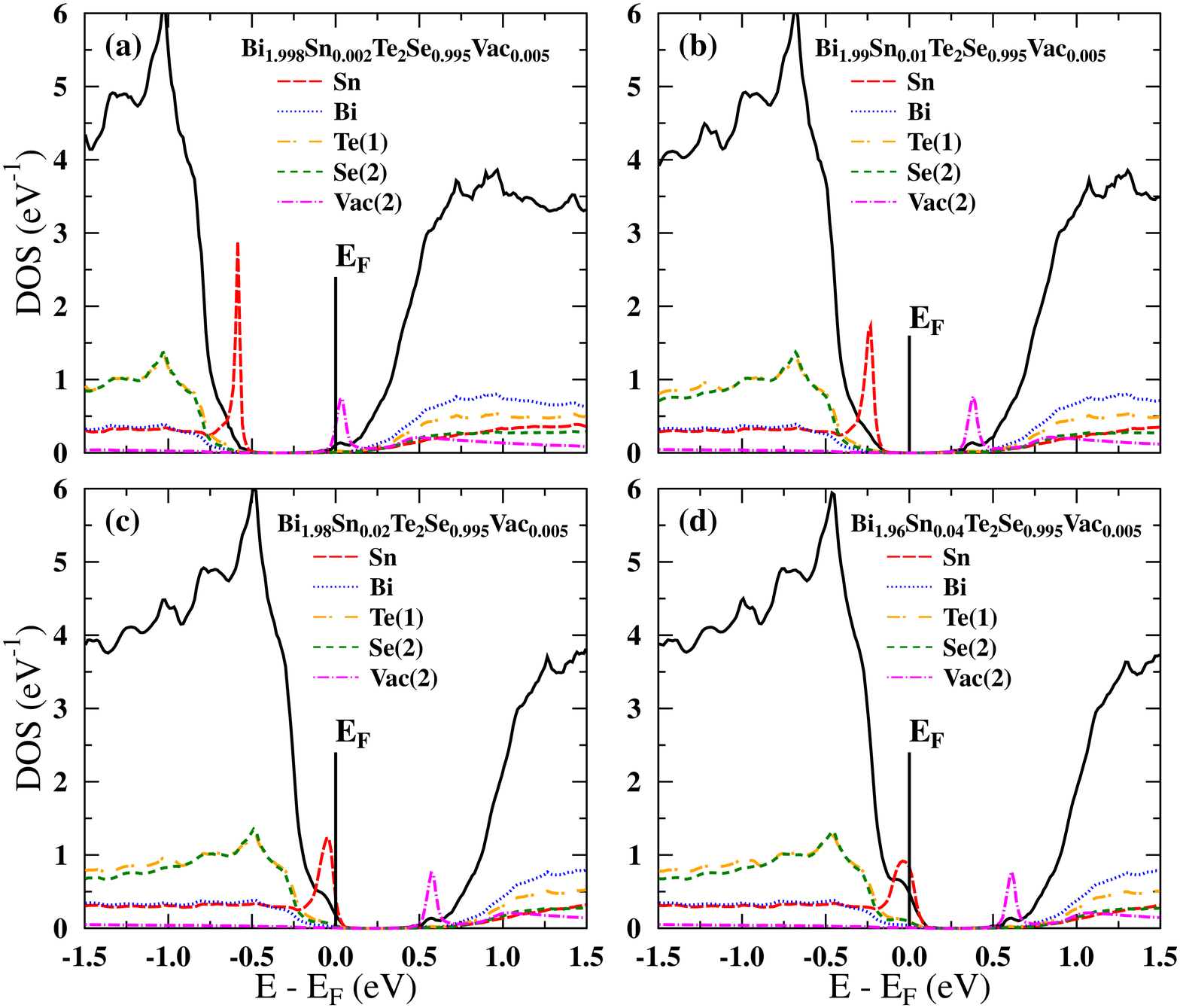}
    \caption{Evolution of DOS and Fermi level position in Sn doped Bi$_2$Te$_2$Se$_{0.995}$, i.e. containing 0.5\% of the 'inner layer' Se(2) vacancies. When Sn concentration is larger than two times the vacancy concentration, $E_F$ starts to penetrate the RL DOS peak (panel c), however for large concentrations (panel d) RL peak becomes rather broad.  
    Results are similar to the 'outer layer' vacancies case, presented in Fig.~\ref{fig:snvactese}.}
  \label{fig:S7}
\end{figure}

\begin{figure}[t]
\includegraphics[width=0.50\textwidth]{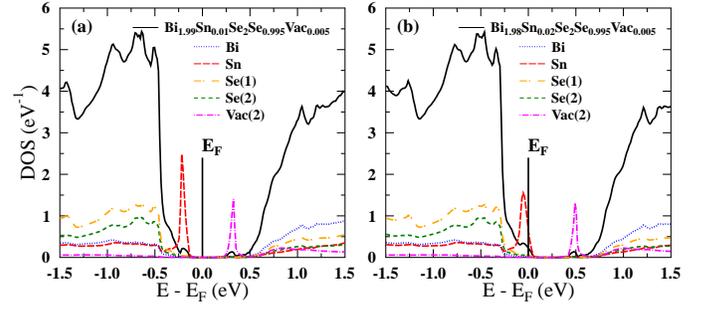}
  \caption{Evolution of DOS and Fermi level position in Sn doped Bi$_2$Se$_{2.995}$ for the 'inner layer' Vac(2) case. $E_F$ behavior and valence DOS are the same, as for the Vac(1) case, presented in Fig.~\ref{fig:S3}.}
  \label{fig:snvacse}
\end{figure}

\begin{figure}[b]
\includegraphics[width=0.50\textwidth]{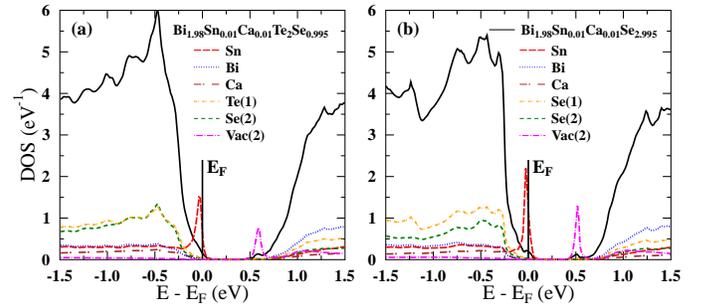}
  \caption{DOS of the doubly, Sn- and Ca-doped materials, containing 0.5\% of the 'inner layer' Se(2) vacancies: (a) Bi$_2$Te$_2$Se and (b) Bi$_2$Se$_3$. Similarly to the Vac(1) case (Fig.~\ref{fig:S5}) our results show that double doping with rigid-band-like acceptor (Ca) should be an effective way of tuning the Fermi position without broadening the resonant level too much.}
  \label{fig:sncavacse}
\end{figure}

Fig.~\ref{fig:vaccamgse} shows evolution of DOS of Bi$_2$Se$_3$ containing vacancies on Se(2), when it is counter-doped using rigid-band-like acceptors Ca and Mg. These two dopants allow to compensate the donor behavior of the Se(2) vacancy in the same way, as if the vacancy was located on Se(1) atoms (cf. Fig.~\ref{fig:S2}).
Similar conclusions are valid for Bi$_2$Te$_2$Se, when Fig.~\ref{fig:vaccamgtese} and Fig.~\ref{fig:S8} are compared.

In Fig.~\ref{fig:S7} we observe, that the evolution of the electronic structure for the Sn doped 
Bi$_2$Te$_2$Se$_{0.995}$, i.e. containing 0.5\% of the 'inner layer' Se(2) vacancies, when the concentration of Sn is increasing, is similar to that observed for the 'outer layer' vacancy case (cf. Fig.~\ref{fig:snvactese}).

Fig.~\ref{fig:snvacse} shows Sn doped Bi$_2$Se$_{2.995}$, with the Se(2) vacancies. DOS and $n$-$p$ crossover, while increasing Sn concentration, is similar as in the Fig.~\ref{fig:S3} for Vac(1).

In Fig.~\ref{fig:sncavacse} $n$-type character of 0.5\% of Vac(2) is compensated using rigid-band-like acceptor (Ca) in the presence of RL at Sn atom in Bi$_2$Te$_2$Se (panel (a)) and Bi$_2$Se$_3$ (panel (b)), without broadening of the Sn resonant level, in the same way as was seen for Vac(1) case in Fig.~\ref{fig:S5}.

\end{document}